\documentclass[referee]{aa}

\usepackage{graphics}
\usepackage{times}
\fontfamily{ptm}
\selectfont

\begin{document}

\thesaurus{03(11.11.1; 11.05.1; 11.03.4; 11.19.6; 11.06.2; 11.07.1)}

\title{Extended stellar kinematics of elliptical galaxies in the Fornax
cluster.\thanks{Based on observations collected at Siding Spring 
Observatory}\fnmsep\thanks{Table 3 is presented in electronic form
only, and is available from the CDS, Strasbourg}}

\author{ A.W. Graham   \inst{1} \and 
         M.M. Colless  \inst{1} \and 
         G. Busarello  \inst{2} \and 
         S. Zaggia     \inst{2} \and 
         G. Longo      \inst{2}    }

\offprints{A.W.\ Graham, ali@mso.anu.edu.au}

\institute{Mount Stromlo and Siding Spring Observatories, Australian
National University, PMB Weston Creek PO, ACT 2611, Australia \and
Osservatorio Astronomico di Capodimonte, via Moiariello 16, I-80131
Napoli, Italy }

\date{Received / Accepted }

\titlerunning{Elliptical galaxies in the Fornax cluster}
\authorrunning{Graham et al.}
\maketitle

\begin{abstract}
We present extended stellar kinematics for a sample of elliptical
galaxies in the Fornax cluster. Out of the 13 galaxies presented here,
five (FCC 119, FCC 136, NGC 1373, NGC 1428, FCC 335) have no previously
published kinematical data. Major-axis velocity dispersion profiles
(VDPs) and rotation curves (RCs) are given for 12 of the galaxies.
A major feature of this data is the spatial extension: for 8 galaxies
the data extends beyond 1 $R_e$, and for 5 it extends beyond 2 $R_e$.
Compared to the previously available data, this corresponds to an
increase in spatial coverage by a factor from 1 to 5.
The present sample represents 86\% of the ellipticals in Fornax brighter
than B$_T$=15 mag.

Five of the ellipticals in the sample turn out to be
ro\-ta\-tio\-nal\-ly-supported systems, having positive rotation
parameter $\log (\frac{V}{\sigma})^*$.
One of these five, and another 3 galaxies from the remaining sample,
display evidence for bar-like kinematics.

The data indicate that the true number of `dynamically hot' stellar
systems, is much lower than previously thought: of the Es in the present
sample only 1/4 are confirmed as `pressure-supported' systems.

The data reveal a host of individual peculiarities, like: wiggles,
strong gradients, and asymmetries in the rotation curve and/or in the
velocity dispersion profile, thus showing that the presence of
kinematically distinct components and/or triaxiality is a common
characteristic of this class of object.

\end{abstract}

\section{Introduction}

The scaling relations between some global properties of ellipticals
(Es), known as the `Fundamental Plane' (FP; Dressler et al. 1987,
Djorgovski \& Davis 1987), is clearly a manifestation of a great
`regularity' in the properties of Es. Although a proof of its
`universality' is still lacking, the FP has been widely applied to
various problems: from the study of peculiar motions (Lynden-Bell et al.
1988, J\o rgensen et al. 1996), to the study of the evolution of M/L
ratios (van Dokkum \& Franx 1996, Kelson et al. 1997), and measurements 
of $q_0$ (Bender et al.\ 1998). 

The `regularity' of the E's properties, shown by the FP, must be connected to 
their gross structural characteristics and to their formation and 
evolutionary mechanisms (Djorgovski et al. 1995 and references therein). To
understand this connection is however a very hard task, especially in
view of the problems posed by the `thinness' of the FP itself 
(Ciotti et al. 1996). 
Apart from the obvious similarities, it is a matter of fact that Es do
not constitute a homogeneous family of galaxies: Es not only have
substantially different photometric and kinematical properties, but they
also host a variety of sub-components, like inner disks,
coun\-ter-ro\-ta\-ting cores, outer rings, dust lanes, etc. (Illingworth
1983, Kormendy \& Djorgovski 1989, Caon et al. 1993, Michard \& Marchal
1994, Capaccioli \& Longo 1994, Scorza \& Bender 1996, Graham et al.\ 1996, 
Graham \& Colless 1997, Busarello et al. 1997, Prugniel \& Simien 1997, 
Carollo et al. 1997, and references therein).

The possibility that many Es have disks that escape photometric
detection (Simien \& de Vaucouleurs 1986, Capaccioli 1987, Rix \& White
1990, Saglia et al. 1997), and the dichotomy between the two families of
boxy and disky Es lead to the possibility of a new classification
scheme, in which galaxies formerly belonging to the same Hubble type are
separated according to isophotal shape, and in which a continuous
sequence exists between Es and S0s (Kormendy \& Bender 1996).
Since the above scenario is manifestly at odds with the `regularity' of the 
FP, the question arises to what extent the `global' quantities entering into
the construction of the FP are representative of the real status of the 
galaxies.

In this respect, an important question is whether the central
velocity dispersion $\sigma_0$ is fully representative of the dynamical
status of a galaxy. There is much evidence that this is not the case:
due to differences in velocity dispersion profiles and anisotropy,
masses derived from $\sigma_0$ can be wrong by a factor of ten (Tonry
1983, Richstone \& Tremaine 1986, Mathews 1988); in presence
of central, kinematically distinct sub-com\-po\-nents, the value of
$\sigma_0$ is likely to be more connected to the sub-com\-po\-nent than to
the galaxy as a whole; Es are dynamically non-ho\-mo\-lo\-gous systems: the
slope of the velocity dispersion profile is correlated to $\sigma_0$ in
such a way that there is a systematic departure of the kinetic energy
from a simple scaling of $\sigma_0$  (Graham \& Colless 1997, Busarello
et al. 1997); finally, the rotation, although not dominant in Es, plays a
non-negligible role (Busarello et al. 1992, Prugniel \&
Simien 1994, 1996, Busarello et al. 1997, D'Onofrio et al. 1997).

To fully understand the complex kinematics of Es, and
to analyze the connection between the individual properties of Es and
the scaling laws, it is essential to have access to kinematical data as
accurate, extended and homogeneous as possible.
One of the objectives of the present work is to obtain homogeneous sets
of data for magnitude-limited samples of elliptical galaxies belonging
to different clusters.

In the present paper we present new stellar kinematical data for 13 
galaxies in the Fornax cluster.

Section 2 describes the observations and the data reduction.
In Sect.\ 3, we give a short description of the major features of each
of the kinematical profiles.
Section 4 provides a comparison of our data with previous
observations, and in Sect.\ 5 we briefly summarize our findings. 
The data themselves are presented in Appendix A and in Table 3 
(available in electronic form from the CDS).

\section{Observations and Reduction}

\subsection{Observations}

Ferguson (1989) lists 340 likely Fornax cluster members within an area
of $\sim$40 sq.deg.\ centered on the cluster.  Of these, 14 are
classified as elliptical galaxies brighter than $B_T$$=$15.0 mag. We
were able to obtain kinematical data for 12 of these galaxies making our
investigation 86\% complete at that limiting magnitude.
In addition, we have obtained kinematical data for FCC 119, (classified as 
a S0 galaxy by Ferguson, and for which we give only the central velocity
dispersion (CVD) and V$_{max}$).

The spectra were obtained over two runs in November and December 1996 
using the blue arm of the Double Beam 
Spectrograph attached to the Australian National University's 2.3 m 
telescope at Siding Spring Observatory.
Dichroic \# 3 was used, providing greater than 95\% transmission 
at all wavelengths between 4000-6000 \AA.
A 1200 {\it l} mm$^{-1}$ grating was used, with a dispersion of 0.555 
\AA $\:$ pixel$^{-1}$ over a range of 1000 \AA $\:$ centered on 5200 \AA.
The CCD used was a SITe chip (1752$\times$532) with 15 $\mu$m pixels.
The spatial scale on the chip was 0.91$\arcsec$ pixel$^{-1}$.
We used a spectrograph slit of 2$\arcsec$ on the sky and of length greater 
than the spatial extent of the CCD.  
The best collimator focus gave a FWHM for the arc lines of 2.7 pixels
or 1.50 \AA, giving a resolution of 86 km s$^{-1}$ at 5200 \AA.
During the observations the seeing varied from 1.5 to 2.5 arcsec .

The observed galaxies are listed in Table~\ref{tab1}, along with their
photometric parameters (taken from Caon et al.\ 1994, we also refer to 
this paper for a detailed photometric description).  
We obtained our estimates of the galaxy major-axis P.A.'s based on careful 
inspection of the position angle (P.A.) profiles (Caon et al.\ 1994).  
Only our estimate for the major-axis P.A.\ of NGC 1419 differed from the
value adopted in Caon et al.\ (1994), where we have adopted 50$^{\circ}$
rather than 65$^{\circ}$.
The number of spectra for each galaxy and the total exposure times used are 
also listed in the table.
The spectra were exposed in 30--45 min blocks.  
A Ne-Ar lamp was observed before and after each exposure for wavelength
calibration and four template stars (of spectral types from G8III to 
K3III)
were observed at the beginning, middle and end of each night. The usual
dome and sky flats were taken, as were bias frames and measurements of
the dark current.

\begin{table*}
\caption[]{Galaxy sample and photometric parameters}
\label{tab1}
\begin{flushleft}
\begin{tabular}{lllrlcrcrcl}
\noalign{\smallskip}
\hline
\noalign{\smallskip}
$\alpha_{1950}$  &  $\delta_{1950}$ & NGC\# & FCC & Type & $B_T$ &$r_e$&$\mu_e$& P.A.\ & \# &Exp.\\
 (h\  m\  s)     &( $\circ$\   $ \prime $\  $ \prime\prime $ ) &   &   &   &  
(mag)&($\arcsec$)&(mag/sq.arcsec.) & $\circ$  &spec.\ &(hr) \\
\noalign{\smallskip}
\hline
\noalign{\smallskip}
 3\ 24\ 35.9 & $-35$\ 53\ 10 & NGC 1336    &  47 & E4    & 13.3 &  30 & 23.6 & 20 & 
2 & $1$ \\
 3\ 26\ 06.1 & $-32$\ 27\ 26 & NGC 1339    &  63 & E4    & 12.7 &  15 & 21.5 &175 & 
4 & $2$ \\
 3\ 31\ 35.6 & $-33$\ 44\ 19 &             & 119 & S0    & 15.1 &  19 & 24.4 & 40& 
3 & $2.25$ \\ 
 3\ 32\ 34.1 & $-35$\ 42\ 39 &             & 136 & dE2   & 14.7 &  26 & 24.7 &170 &  
5 & $2$ \\
 3\ 33\ 03.3 & $-35$\ 20\ 06 & NGC 1373    & 143 & E3    & 14.2 &  11 & 22.4 &140 & 
7 & $3.5$\\
 3\ 33\ 21.1 & $-35$\ 23\ 29 & NGC 1374    & 147 & E0    & 11.9 &  26 & 22.3 &120 & 
2 & $1$ \\
 3\ 34\ 08.7 & $-35$\ 36\ 22 & NGC 1379    & 161 & E0    & 12.0 &  24 & 22.0 & 07 & 
2 & $0.75$ \\
 3\ 36\ 34.2 & $-35$\ 36\ 46 & NGC 1399    & 213 & E0    & 10.0 & 134 & 24.2 &112 & 
5 & $3$ \\
 3\ 36\ 57.3 & $-35$\ 45\ 17 & NGC 1404    & 219 & E2    & 10.9 &  26 & 21.2 &160 & 
3 & $2.5$ \\
 3\ 38\ 50.3 & $-37$\ 40\ 09 & NGC 1419    & 249 & E0    & 13.6 &   9 & 21.7 & 50 & 
2 & $1$ \\
 3\ 40\ 24.6 & $-35$\ 33\ 06 & NGC 1427    & 276 & E4    & 11.8 &  39 & 22.8 & 77 & 
2 & $1$ \\
 3\ 40\ 27.5 & $-35$\ 18\ 44 & NGC 1428    & 277 & E5    & 13.8 &  12 & 21.9 & 115&
 7 & $3.25$ \\
 3\ 48\ 43.8 & $-36$\ 03\ 29 &             & 335 & E     & 14.4 &  19 & 23.5 & 45 & 
4 & $2.25$\\
\noalign{\smallskip}
\hline
\end{tabular}
\end{flushleft}
The columns respectively list the R.A.\ and Dec.\ of each galaxy, the
NGC number (where available) and Fornax Cluster Catalogue number (FCC; Ferguson
1989), morphological type according to Ferguson (1989), total apparent
blue magnitude, effective half-light radius and associated blue surface
brightness at that radius and major-axis position angle (measured E from N). 
The number of spectra and the total exposure time for each object are also 
given.  The photometric parameters are taken from Caon et al.\ (1994).
\end{table*}

\subsection{Data Reduction}

The data reduction included, besides the usual CCD cosmetics, noise
removal along the spatial direction by adaptive filtering techniques
(Richter et al. 1992). 
After wavelength calibration and sky subtraction, the spectra
were normalized to the con\-tinu\-um, obtained by fitting with a 6th order
polynomial. 
The reliability of the sky subtraction was iteratively tested by 
comparing the brightness profiles of the sky-subtracted spectra to the 
profiles from Caon et al. (1994). The slit function, as derived from 
twilight spectra, varied by at most 0.5\%, so that no further correction 
was needed.
The low-frequency residual variations were reduced by filtering in
Fourier space. Finally, the spectra were processed by the Fourier
Correlation Quotient (FCQ) technique (Bender 1990). All four template
stars were used, and the relative results were compared to test their
consistency: as expected, no relevant difference was detected (see
Bender 1990).
In order to test the reliability of the observed features, we compared
the velocity dispersion profiles (VDPs) and the rotation curves (RCs) 
obtained by adding the different spectra of each
individual galaxy before and after the final stage of processing (the
FCQ), obtaining fully consistent results.
As a further test we also reduced the data without filtering and, 
although the data were obviously more noisy in the fainter regions, the 
observed features were still present.
A lower limit of $\sim$35 km s$^{-1}$ in the measurable velocity
dispersion, due to the instrumental setup, was also verified on the
template stars.
The uncertainties in the data are those derived by the FCQ procedures,
basically arising from the fit of the broadening function, and are
explained in Bender (1990).

\section{Results}

The figures in Appendix A and Table 3 (available in electronic form)
present the RCs and VDPs. 
The major-axis P.A. points from left to right in the figures, that is,
the right hand side in the plots corresponds to the easterly half of 
the galaxies major-axis. 
In addition, in Table 2 we list the values of the CVD and of the maximum 
observed rotation velocity $V_{max}$.

The major kinematical features of each galaxy are outlined in the following.
The term `kinematically distinct' is used in the present Section in a 
descriptive sense only, without attaching to it any particular 
interpretation (see Sect. 5). 
By `inner (or central) $x\arcsec$', we refer to a radius (not a 
diameter).

\vspace{3mm}
\noindent 
{\bf NGC 1336}

The VDP dips $\sim$5 km s$^{-1}$ over the central 2$\arcsec$ and contains 
additional oscillations at larger radii, having amplitudes of 
10 km s$^{-1}$. Smoothing these out, the VDP shows negligible signs of 
decreasing until a radius of 18$\arcsec$, remaining roughly flat at 
$\sim$100 km s$^{-1}$.  
The RC is symmetric, being approximately flat over the inner
$\sim$5$\arcsec$, where a strong isophotal twist is known to take place (up to
60$^{\circ}$; Caon et al.\ 1994).
Beyond this radius, the RC displays a wave-like appearance about a zero
mean rotational velocity, with an amplitude of 20 km s$^{-1}$.  Such
kinematics is indicative of the presence of a bar (Sparke \& Sellwood
1987; Bettoni 1989, and references therein).
Concerning this point, it is also worth noting that the normalized
Fourier coefficient $a_4$ is high over the inner 10$\arcsec$ (Caon et
al.\ 1994).

\vspace{3mm}
\noindent 
{\bf NGC 1339}

The RC is well traced out to 2 R$_e$, having a steep gradient of 200 km
s$^{-1}$ kpc$^{-1}$ (assuming a distance of 18 Mpc) over the inner
5$\arcsec$ radius.  The E and W side of the RC largely flatten out after
10$\arcsec$.  The VDP decreases linearly from 175 km s$^{-1}$ to 
100 km s$^{-1}$ at 10$\arcsec$ on both sides. 
After 20$\arcsec$, both sides of the VDP increase some 50 km s$^{-1}$, reaching
the outer data point (30 $\arcsec$). 
%It then remains flat on the E side while it drops to 70 km s$^{-1}$ on the W side.
%The Fourier coefficient $a_4$ (Caon et al.\ 1994) also shows a marked 
%disturbance at this radius.

\vspace{3mm}
\noindent
{\bf FCC 119}

Catalogued as a S0 galaxy by Ferguson (1989), we only provide the
central velocity dispersion $\sigma_0$=51$\pm$10 km s$^{-1}$ and maximum
observed rotational velocity $V_{max}$=27$\pm$19 km s$^{-1}$ for this
galaxy, the faintest of our sample.

\vspace{3mm}
\noindent
{\bf FCC 136}

This faint galaxy is classified as dE2 by Ferguson (1989) and as SAB0 in 
RC3.
The VDP is poorly defined around 40 km s$^{-1}$, close to our measurement
limit. 
There is marginal evidence for rotation within the wavy RC, which 
increases to 15 km s$^{-1}$ by the outer data points (15$\arcsec$).
The shape of the RC, together with the fact that the inner isophotes
($<$4$\arcsec$) twist through 30$^{\circ}$, while the normalized Fourier
coefficient $a_4$ is high (Caon et al.\ 1994) supports the RC3 
classification as a barred galaxy.  

\vspace{3mm}
\noindent
{\bf NGC 1373}

The symmetric RC reveals a small degree of rotation, 
peaking at 15 km s$^{-1}$.
The inner 10-15$\arcsec$ of the VDP has a wave-like appearance,
with a 5 km s$^{-1}$ wave amplitude producing no obvious, and several local,
maxima around 80 km s$^{-1}$.  
The VDP is asymmetric: the E side is flat out to 14$\arcsec$ from the 
center and then decreases to $\sim$30 km s$^{-1}$ at 20$\arcsec$, while the 
W side slowly decreases starting from 6$\arcsec$, to reach a minimum of 
40 km s$^{-1}$ at 24$\arcsec$.

The overall shape of the VDP and the presence of strong isophotal twisting 
(Caon et al. 1994) are, in this case, also indicative of the presence 
of a bar.

\vspace{3mm}
\noindent
{\bf NGC 1374}

The inner gradient of the rotation velocity is very steep (250 km s$^{-1}$ 
kpc$^{-1}$). After reaching a first maximum at 2$\arcsec$ from the center, the 
RC remains approximately flat on both sides, until reaching a second 
maximum at 30--35$\arcsec$. 
The asymmetries of the RC correspond to the analogous behavior of the 
VDP which, after an initial strong gradient, and after reaching a plateau 
of 145 km s$^{-1}$ at 7--10$\arcsec$, shows different slopes on the two sides.  
The VDP is approximately flat, at 120 km s$^{-1}$, on the E side in the 
range 15-35$\arcsec$ before dropping to a minimum of 70 km s$^{-1}$ 
at the same position (35$\arcsec$) of the maximum of the RC.  However, 
the W side of the VDP shows evidence for increasing over the radial range 
corresponding to the second rise in the W side of the RC.  

\vspace{3mm}
\noindent
{\bf NGC 1379}

The kinematics of this galaxy, although rather symmetric in its overall
appearance, presents a number of remarkable small-scale asymmetries.
The series of minima and maxima in both the RC and the VDP are clearly
anti-correlated, the maxima of the VDP corresponding to minima in the
RC and vice-versa.
There is evidence for a small (5$\arcsec$) inner disk, coming from the
steepest part of the RC, clearly connected to the inflexion in both
sides of the VDP at this radius.
There is also marginal evidence for a central distinct component 
corresponding to the inflexion of the RC.

\vspace{3mm}
\noindent
{\bf NGC 1399}

This is the largest Fornax galaxy, at the center of the cluster.
Its dominant status is confirmed by its large velocity dispersion (notice 
however that the peak value of $\sigma$ is underestimated by our 
data because of the relatively poor seeing, see Sect. 4). 
The VDP quickly declines to 
250 km s$^{-1}$ by a radius of 10$\arcsec$, to remain approximately flat 
to the outer data points.
The RC presents strong evidence for a kinematically distinct inner component,
with an inflexion point evident over the central 4$\arcsec$. 
The W side of the RC reaches a maximum value of 30 km s$^{-1}$ at 
15$\arcsec$, where it then steadily falls to zero by 55$\arcsec$.
It then changes sign, increases to 10 km s$^{-1}$ at 60$\arcsec$ and 
then decreases again.
The wiggles of the RC in the range 40-70$\arcsec$ are associated 
to wiggles in the VDP in the same range and with the same amplitude.
The E side\ of the RC appears identical to the W side\ until 
reaching 30$\arcsec$ where it increases to 50 km s$^{-1}$ 
to then stagger back down to a rotational speed of 20 km s$^{-1}$ at
the outer data point.  The E side\ of the VDP resembles the W side\ until
at 55$\arcsec$ where it decreases by 40 km s$^{-1}$ over the following
5$\arcsec$ and then starts climbing, still increasing at the outer data 
point.  

\vspace{3mm}
\noindent
{\bf NGC 1404}

The RC has a steep gradient of 200 km s$^{-1}$ kpc$^{-1}$ inside 10 
$''$, with the exception of the two inner arcseconds, where
the RC remarkably flattens. 
At $r$$\sim$$10$$''$ and $r$$\sim$$22$$''$ there are two local maxima around
$V$$\sim$$100$ km s$^{-1}$, which are symmetric with respect to the center.
The rotation velocity starts increasing again beyond $r$=$30$$''$,
reaching $\sim$$140$ km s$^{-1}$ at the last observed point: the RC seems
actually to be increasing even beyond 3 effective radii.

In the central region the VDP dips by 15 km s$^{-1}$ from the two local 
maxima situated at $\pm 2''$ from the center. The velocity dispersion 
then decreases at a nearly constant rate in the inner 20$''$, until 
reaching a plateau around 200 km s$^{-1}$.
In the range from 40$''$ to 50$''$ the VDP decreases before increasing
to a local maximum of 180 km s$^{-1}$ around 1 arcmin from the center.

\vspace{3mm}
\noindent
{\bf NGC 1419}

This galaxy has a very similar RC to NGC 1336, with $\sim$15 km s$^{-1}$
wiggles observed out to 2 $R_e$. 
The VDP is largely flat (at $\sim$120 km s$^{-1}$) 
over the inner 9$\arcsec$(=1$R_e$), then it steadily decreases to 
$\sim$70 km s$^{-1}$ reaching the outer radius limit of the data.
Like NGC 1336, this galaxy also has a very strong 
isophotal twist ($\sim$100$^{\circ}$) over the inner region, where the RC is
observed to remain flat.  
Also in this case we thus find evidence for a possible bar-like 
structure.

\vspace{3mm}
\noindent
{\bf NGC 1427}

This galaxy has a wiggly VDP which peaks around 180 km s$^{-1}$ and falls
to 140 km s$^{-1}$ at the data boundary.  The CVD is $\sim$13 km s$^{-1}$ 
below the peak value which occurs 4$\arcsec$ W from the center.
The RC, after reaching a first maximum of $\sim$25 km s$^{-1}$ at 5$\arcsec$,
shows a series of secondary maxima on both sides, that are 
anti-correlated to the wiggles in the VDP.
There is also marginal evidence for an inflexion in the center which, 
if real, and if related to the dip in the VDP could constitute kinematical 
evidence for the central faint disk detected by Carollo et al. (1997, 
see also Forbes et al. 1995).

\vspace{3mm}
\noindent
{\bf NGC 1428}

The presence of a star superimposed on the nucleus of this galaxy 
prevents one from obtaining kinematical data in the central 4$\arcsec$. 
In the observable region, the rotation velocity steadily increases, 
reaching $\sim$70 km s$^{-1}$ at 2.7 $a_e$ on both sides.
Small amplitude waves are present around 15$''$ from the center in 
both the RC and the VDP. 
A general, though weak, asymmetry is observed in both profiles: on the E 
side the velocity dispersion is lower and the rotation velocity is 
higher than on the W side. 
The increase of the VDP beyond 20$''$ in the E side is, on the other
hand, not real, as can be inferred from the large uncertainties.

\vspace{3mm}
\noindent
{\bf FCC 335}

This is a faint galaxy, with a roughly flat VDP of 40 km s$^{-1}$, close
to the lower limit of our measurements.
There is marginal evidence for an increase on the W side of the VDP of 
$\sim$10 km s$^{-1}$ around 13$\arcsec$.  This is associated with a decrease 
of the RC by 15 km s$^{-1}$ over the same interval.  The RC itself peaks at
a maximum rotational velocity of $\sim$25 km s$^{-1}$ at 
10$\arcsec$=0.5 $R_e$, where the W side\ then decreases its 
rotation and the E side\ remains flat to the boundary of our data. 

\begin{table*}
\caption[]{Summary of the kinematical properties.}
\label{tab2}
\begin{flushleft}
\begin{tabular}{lrrrcccr}
\noalign{\smallskip}
\hline
\noalign{\smallskip}
Ident. & $V_{hel}$ &$\sigma_0$   & $V_{max}$   & $r_{max}$ &
$\frac{r_{max}}{a_e}$ & $\sigma_0$ (D95) & log$(\frac{V}{\sigma})^*$ \\
       & km s$^{-1}$ & km s$^{-1}$ & km s$^{-1}$ & ($\arcsec$)    &          
            & km s$^{-1}$ &   \\
\noalign{\smallskip}
\hline
\noalign{\smallskip}
NGC 1336          & 1418$\pm$3  & 108$\pm$\phantom{0}4 &  17$\pm$10  &
25 & 0.8 & 115$\pm$\phantom{0}6  & -0.57 \\
NGC 1339          & 1392$\pm$3  & 174$\pm$\phantom{0}6 &
134$\pm$\phantom{0}8  & 30 & 2.0 & 190$\pm$\phantom{0}6  &  0.08 \\
FCC \phantom{3}119& 1374$\pm$7  &  51$\pm$10 &  27$\pm$\phantom{0}9  &
....... &  ....... & 50         &  0.12 \\
FCC \phantom{3}136& 1205$\pm$1  &  39$\pm$\phantom{0}2 & 
16$\pm$\phantom{0}5  & 15 & 0.6 &  37         & -0.11 \\
NGC 1373          & 1334$\pm$2  &  79$\pm$\phantom{0}1 &  14$\pm$\phantom{0}7  
& 24 & 2.2 &  80         & -0.49 \\
NGC 1374          & 1294$\pm$2  & 187$\pm$\phantom{0}5 &  96$\pm$\phantom{0}9  
& 40 & 1.5 & 225$\pm$10  &  0.22 \\
NGC 1379          & 1324$\pm$2  & 135$\pm$\phantom{0}4 &  30$\pm$\phantom{0}5  
& 60 & 2.5 & 140$\pm$10  &  0.11 \\
NGC 1399          & 1425$\pm$4  & 353$\pm$19 &  51$\pm$\phantom{0}5  & 70 & 0.5 
& 420$\pm$27  & -0.36 \\
NGC 1404          & 1947$\pm$4  & 241$\pm$11 & 140$\pm$12  & 80 & 3.1 &
260$\pm$11  &  0.22 \\
NGC 1419          & 2133$\pm$2  & 118$\pm$\phantom{0}5 & 
14$\pm$\phantom{0}8  & 18 & 2.0 & 125$\pm$\phantom{0}8  &  ($\infty$) \\
NGC 1427          & 1388$\pm$3  & 170$\pm$\phantom{0}7 & 
27$\pm$\phantom{0}2  & 40 & 1.0 & 180$\pm$\phantom{0}5  & -0.62 \\
NGC 1428          & 1640$\pm$8  & .......  &  83$\pm$\phantom{0}8  & 33 & 2.7 
&....... & 
.......  \\
FCC \phantom{3}335& 1430$\pm$2  &  43$\pm$\phantom{0}3 &  24$\pm$\phantom{0}4  
& 15 & 0.8 &  41         & -0.14 \\
\noalign{\smallskip}
\hline
\end{tabular}
\end{flushleft}
The following quantities are listed for each galaxy: heliocentric
velocity (Col. 2); central velocity dispersion (Col. 3); maximum
observed rotation velocity (Col. 4); radial range of the kinematical
data in arcseconds and in units of $a_e$ (columns 5 and 6 respectively),
except for the S0 galaxy FCC 199 (see text); central velocity dispersion
from D95 and the `corrected' value (those without error, see text). The
last column gives the `rotation parameter' $\log (\frac{V}{\sigma})^*$
which measures the relevance of the rotational support (notice that this
value is virtually infinite for NGC 1419, for which $\epsilon_e=0.00$
(Caon et al. 1994)). For NGC 1428, $V_{hel}$ has been derived by
interpolation of the RC in the central region. For FCC 119 $V_{max}$ is
an upper limit.

\end{table*}

\section{Comparison with previous data}

Out of the 13 galaxies in the present sample, five (FCC 119, FCC 136,
NGC 1373, NGC 1428 and FCC 335) have no previously published kinematic data,
while the others have been studied by different groups. Figures 1 and 2 
in Appendix B show our data compared to the previous studies.
We summarize here how our data compare with the other groups results.

Three of our sample galaxies are in common with the study by Franx et al.
(1989; hereafter FIH).
Our RC of NGC 1379 is in close agreement, while our VDP {\it has a
shallower gradient compared to that by FIH}, but agrees with the CVD
measure of $\sim$135 km s$^{-1}$.  It is interesting to note that 
the minor-axis VDP of FIH is instead fully consistent both with ours and 
with the VDP of D'Onofrio et al. 1995.
Both the VDP and the RC of NGC 1399 and NGC 1404 agree with those by FIH.
Bicknell et al.\ (1989) and Stiavelli et al. (1993) presented
kinematical profiles of NGC 1399 at P.A.=84$^{\circ}$. 
Our VDP is consistent with their data, while the comparison with the RC
by Bicknell et al. (1989) is made difficult by the larger uncertainties
in their results (their RC is actually consistent with any RC not
exceeding 30 km s$^{-1}$).
Notice, however, that there is a difference of 28$^{\circ}$ in 
P.A.
We are also consistent with the kinematical profiles by Van der Marel \&
Franx (1993) for NGC 1374, the only galaxy from their sample in 
common with ours.  

The study with the largest number (8) of galaxies in common with us, is that
by D'Onofrio et al.\ (1995; hereafter D95, see also Longo et al. 1994).  
Their kinematical profiles are folded around the photometric 
centers.
In general, the agreement between the two sets of data is good,
with only some of discrepancies in the profiles worth mentioning 
(the CVDs will be addressed below).  
The first discrepancy is the up-turn in the VDP of NGC 1339, apparent in
the data of D95 at 4-6 arcsec.  This difference cannot be attributed to
the folding process used by D95 and it is at odds with our data which
give no indication of its presence.
Another discrepancy is the hump in the RC of NGC 1427 over the 
inner 6 arcsec.  D95 show it to reach a maximum rotational speed of 
$\sim$45 km s$^{-1}$, whereas our data suggests it does not reach
any more than $\sim$25-30 km s$^{-1}$.
The data for NGC 1404 are in agreement in the inner 10$''$, even if the 
central dip in the VDP is less pronounced in the D95 data.
The comparison is difficult beyond that radius, due to the heavy 
smoothing of the D95 data, that eventually lead to a wrong negative 
gradient of the RC.
Finally, the RC of NGC 1339 in D95 is steeper than ours in the inner 
5$''$, while the opposite happens for NGC 1374, where D95 fail do detect 
the maximum at 3$''$ from the center.

The comparison of the CVDs needs some prior comments.
Generally, our CVD measurements agree with the values given in the
above-mentioned works, and with those listed by McElroy (1995), with the
exception of NGC 1399, for which McElroy gives a value of 308 km
s$^{-1}$.
D95 give 420$\pm$27 km s$^{-1}$, that is still under the 430 km s$^{-1}$
derived by Stiavelli et al. (1993) using core resolution techniques
applied to observations performed under 0.7$\arcsec$ seeing. Our CVDs
are systematically lower than those by D95, that were obtained under
better seeing conditions (private communication).
On the other hand, since the smoothing effect of the seeing is stronger
for the steeper VDPs, and since the gradient of the VDP is correlated to
the CVD (Busarello et al. 1997), we expect the lower CVDs to be less
affected by the smoothing. The two sets of data are related by: 
$\log(\sigma_0^{\tiny D95})=1.10(\pm.04)\times \log(\sigma_0^{\tiny
this\ work})-0.17(\pm.1)$ (see Fig. 1).
We can then use this relation to estimate the `corrections' to the CVDs
for the other objects of our sample. These `corrected' CVDs are also
listed in Table 2.

\begin{figure}
\resizebox{\hsize}{!}{\includegraphics{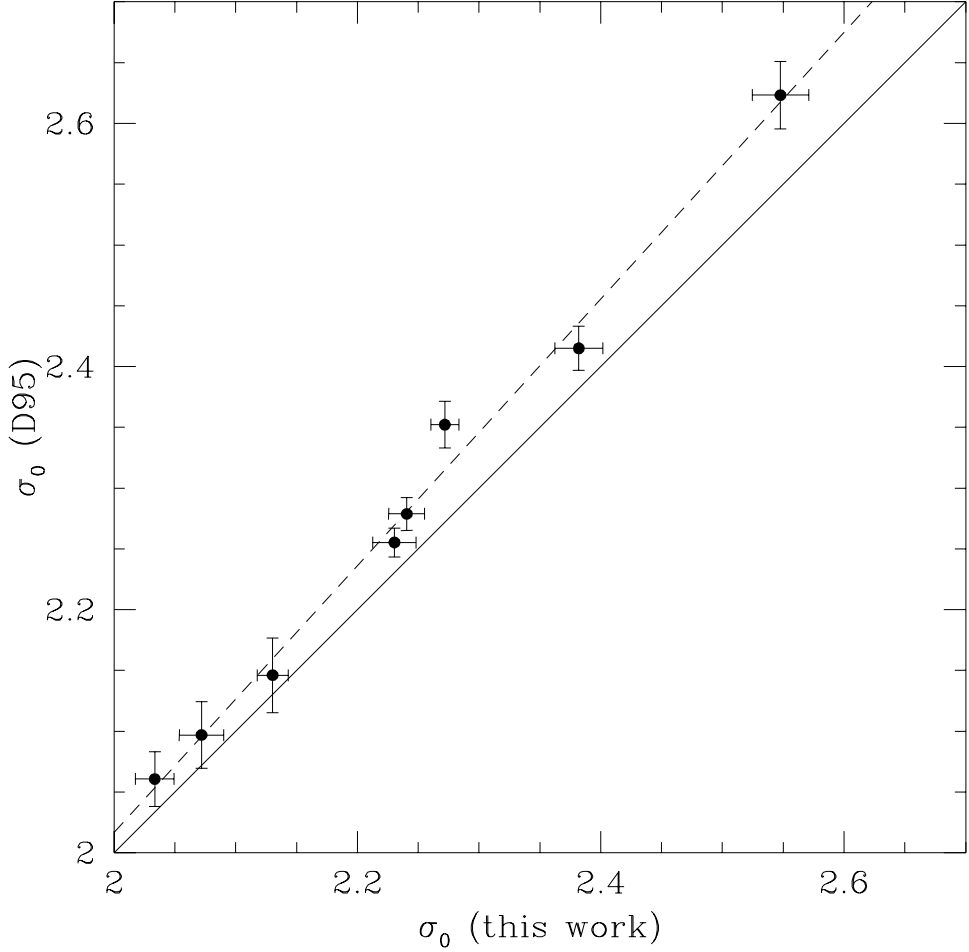}}
\caption{Comparison of the CVDs in this work to those by D'Onofrio et 
al. 1995. The dashed line is the least square fits between the two
samples ($\log(\sigma_0^{\tiny D95})=1.10(\pm.04)\times
\log(\sigma_0^{\tiny this\ work})-0.17(\pm.1)$), while the continuous
line represents the equality between the two quantities. The difference
between the two sets is accounted for by the difference in the seeing
conditions.}
\label{figure1}
\end{figure}

\section{Conclusions}

Of the 12 ellipticals studied here, five (NGC 1339, NGC 1374, NGC 1379, 
NGC 1404 and NGC 1419) are rotationally supported systems, having positive
rotation parameter (log($\frac{V}{\sigma}$)*$>$0).
The interest in this result is strengthened by the fact that four of
these galaxies are E0/E1 systems, the only remaining round system in the 
sample being NGC 1399.
NGC 1374, NGC 1379 and NGC 1404 are most probably S0 galaxies (due to
the overall constancy of their RCs; see also Fisher 1997, and the case
of NGC 3379 presented in Statler et al. 1997), although a more complex
structure has to be invoked to explain their kinematical features.
The RCs of NGC 1374 and NGC 1404, in particular, may reveal the presence
of a double-disk structure (cf. Seifert \& Scorza 1996). 
The rotation parameter of NGC 1339 is consistent with that of an
isotropic oblate model, but the shape and the asymmetry of both the VDP
and the RC indicate a significant departure from a `simple' isotropic
rotator.

Four objects (NGC 1336, FCC 136, NGC 1373 and NGC 1419) are most
probably SB0 galaxies: NGC 1336 and NGC 1419, in particular, have
kinematical profiles identical to the SB0 galaxy IC 456 (Bettoni 1989,
see also Galletta 1996).
This conclusion is also strongly supported by the photometric
properties, all these galaxies having strong isophotal twisting (see the 
photometric profiles in Caon et al. 1994).

The three remaining galaxies are definitely `dynamically hot' 
systems. 
FCC 335 is a faint galaxy (M$_B$=$-$16.8 for H$_0$=75 km s$^{-1}$
Mpc$^{-1}$) with low surface brightness ($\mu_e$=23.5 B mag/arcsec$^2$), low
velocity dispersion ($\sigma_0$= 43 km s$^{-1}$) and low rotation
(V$_{max}$= 24 km s$^{-1}$).
NGC 1399 and NGC 1427 are instead bright galaxies , with high (and
roughly constant) velocity dispersions, and relatively slow rotation:
this is exactly what one expects from `true' giant ellipticals. 
For both objects the present data show complex kinematics: the
already-known kinematically distinct core in NGC 1399 plus a second clear
kinematically distinct component visible at 10$\arcsec$ from the
center. Both from the VDP and from the RC of NGC 1427 there is evidence
for kinematically distinct components at the inner 2$\arcsec$ and
inside 10$\arcsec$ from the center; NGC 1427 shows also anti-correlated wiggles
in both the RC and the VDP.

In summary: it seems that more than half of the galaxies in the present
sample are, at least from the kinematical point of view, misclassified
S(A-B)0s. 
Morphological classification by visual inspection of images, apparently 
results in a dramatic over-estimate of the true number of `dynamically 
hot' stellar systems.  The catalog of Ferguson (1989) lists 58 Fornax 
galaxies with $B_T$$<$15.0 mag, from the 14 listed as having 
morphological type E, only 3 turn out to be true elliptical galaxies 
in the classical sense of the word; indicating that these objects are 
in fact quite rare.
% XXX   Is this okay ???
Most of the objects, moreover, show complex kinematical profiles, like
kinematically distinct components, wiggles, and asymmetries.

The presence of (dynamically decoupled) sub-com\-po\-nents is one of the 
possible explanations of such complex kinematics.
Complicated velocity fields may arise as well from triaxial systems due
to competing contributions of different families of orbits (Statler
1991a,b; Statler et al. 1997, and references therein), like, for
instance, the possible presence of a substantial fraction of
retrograde orbits (Wozniak \& Pfenninger 1997; compare their Fig. 5(c) to
some of our RCs).

A concluding remark is in order: all the data presented here show very
complicated kinematics, which is only in part accounted for by
misclassification of S0s into Es. It seems, instead, conceivable that
complex kinematics is a common, and dynamically stable, characteristic
of elliptical galaxies.

\begin{acknowledgements}
We are grateful to F. Bertola, D. Burstein, M. Capaccioli, K. Freeman,
Ph. Prugniel, F. Simien and L. Sparke for helpful discussions and
suggestions. We are also grateful to the staff of MSSSO and to K.
McKenzie for the telescope assistance. We thank the anonymous referee 
whose comments helped us express some points more clearly. 
\end{acknowledgements}

\appendix
\section{Kinematic profiles}

Velocity dispersion profiles (VDPs) and rotation curves (RCs) for the 
12 galaxies.
The major-axis P.A. points from left to right in the figures, that is,
the right hand side in the plots corresponds to the easterly half of the 
galaxies major-axis.

\begin{figure}
\resizebox{\hsize}{!}{\includegraphics{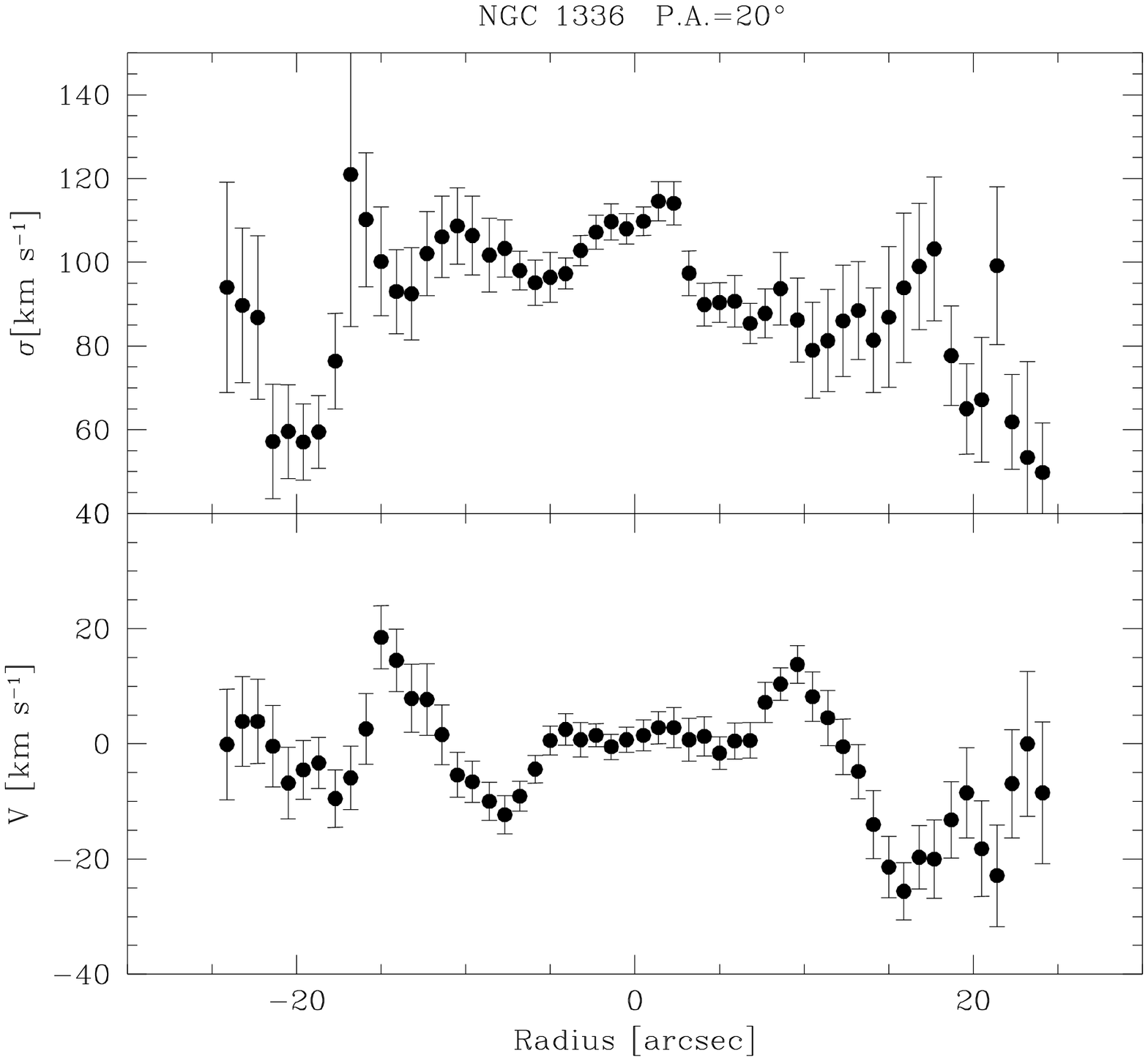}}
\caption{NGC~1336}
\label{n1336}
\end{figure}

\begin{figure}
\resizebox{\hsize}{!}{\includegraphics{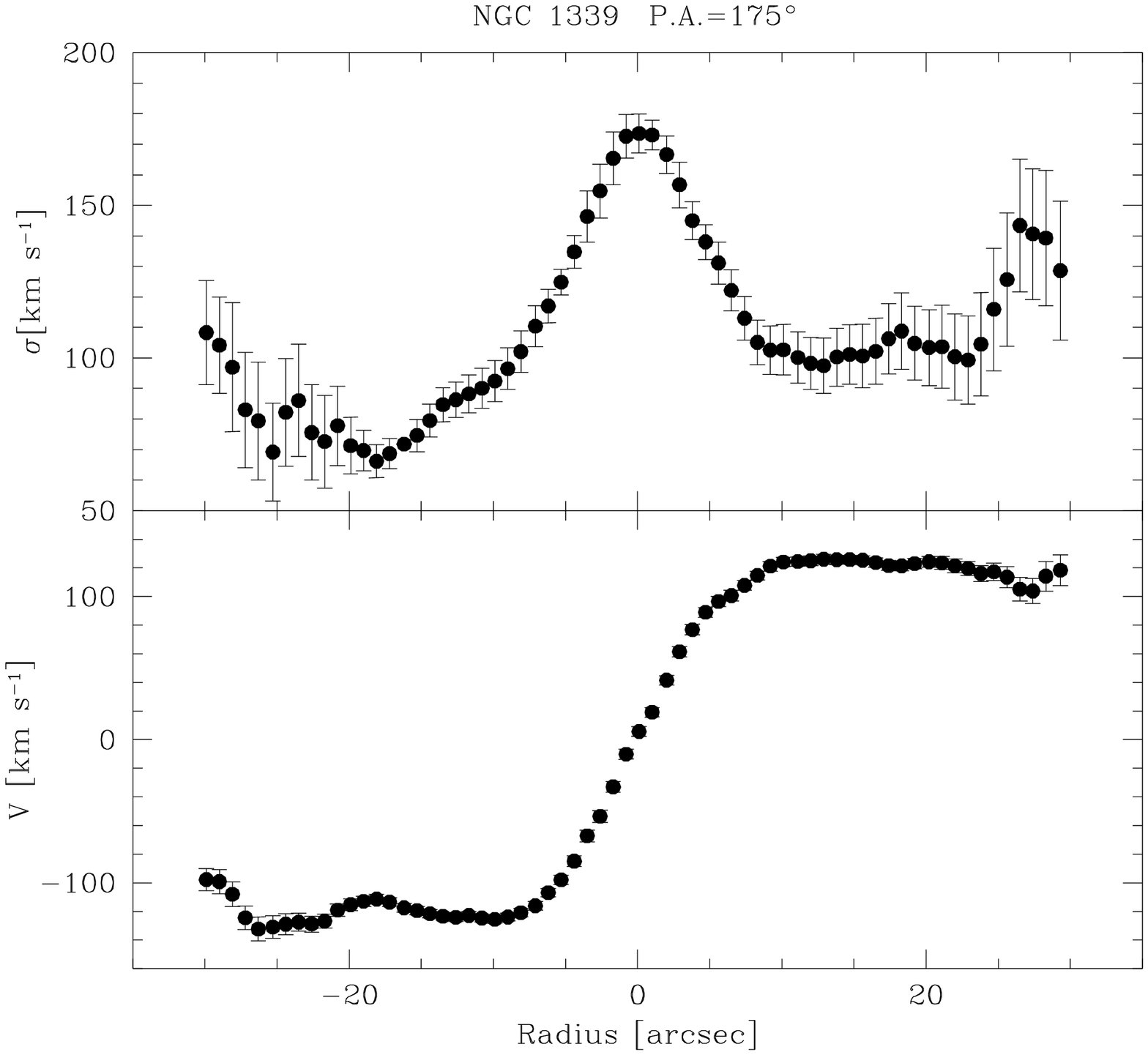}}
\caption{NGC~1339}
\label{n1339}
\end{figure}

%\newpage
%\pagebreak

\begin{figure}
\resizebox{\hsize}{!}{\includegraphics{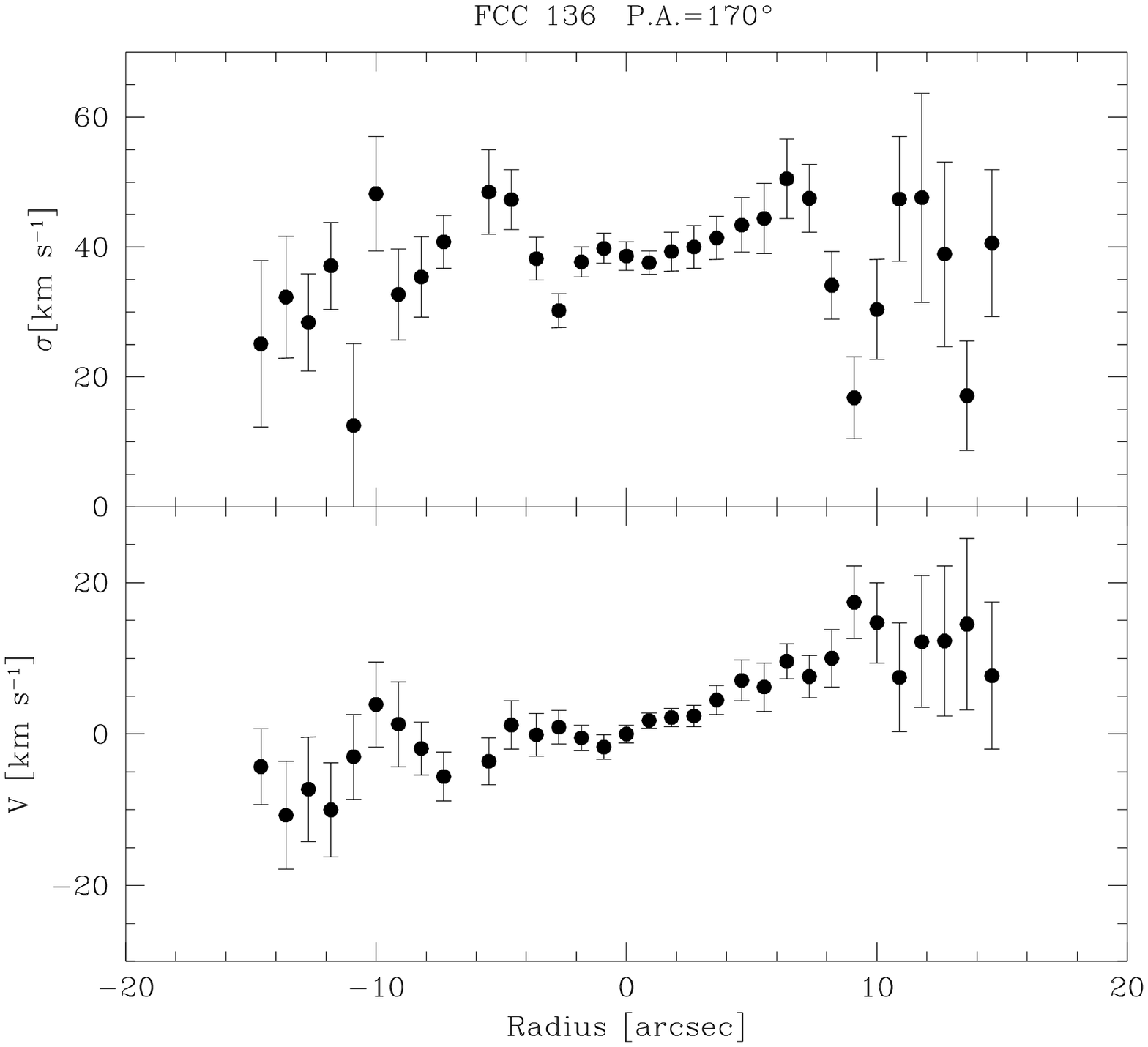}}
\caption{FCC~136}
\label{f136}
\end{figure}

\begin{figure}
\resizebox{\hsize}{!}{\includegraphics{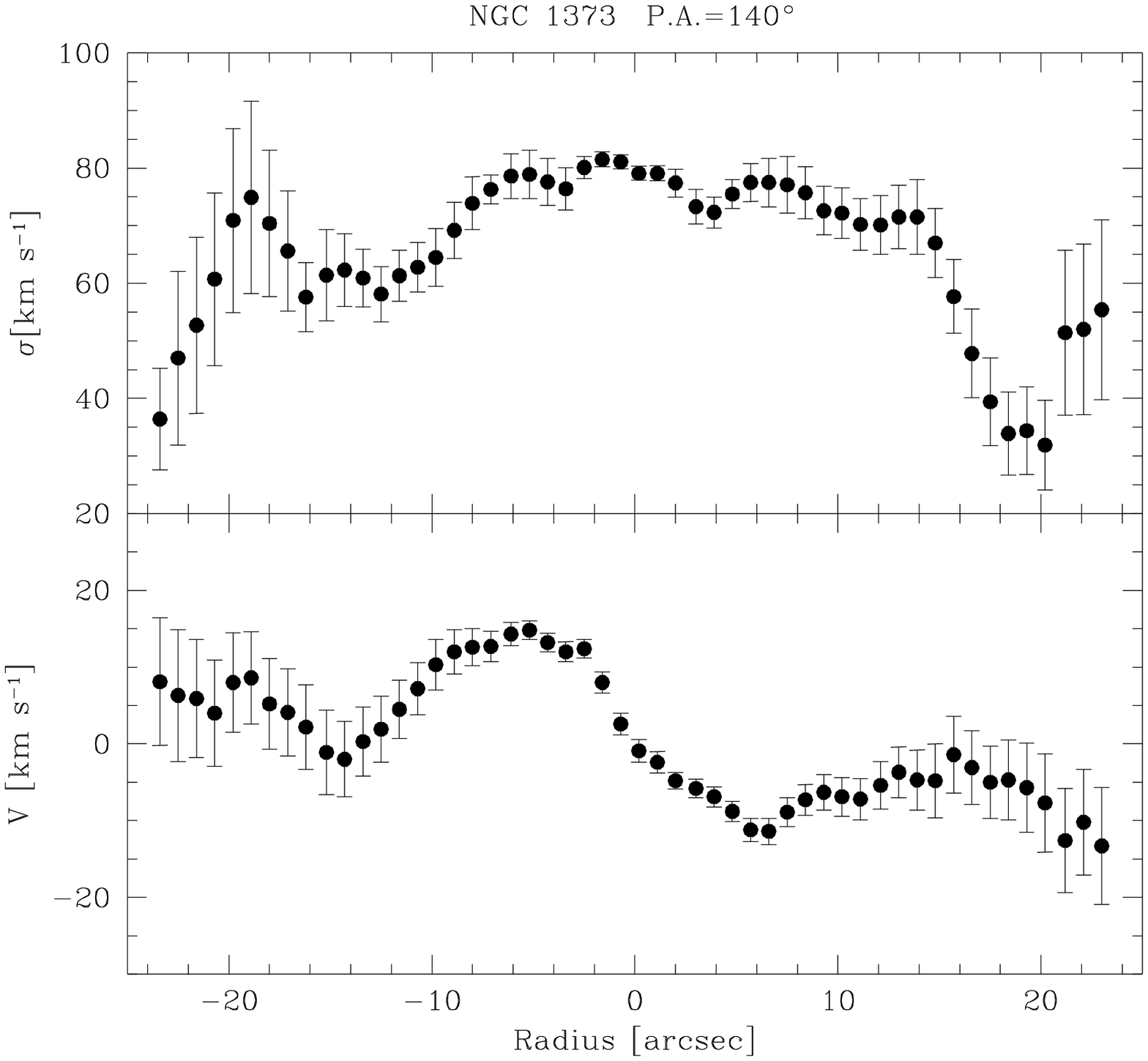}}
\caption{NGC~1373}
\label{n1373}
\end{figure}

\begin{figure}
\resizebox{\hsize}{!}{\includegraphics{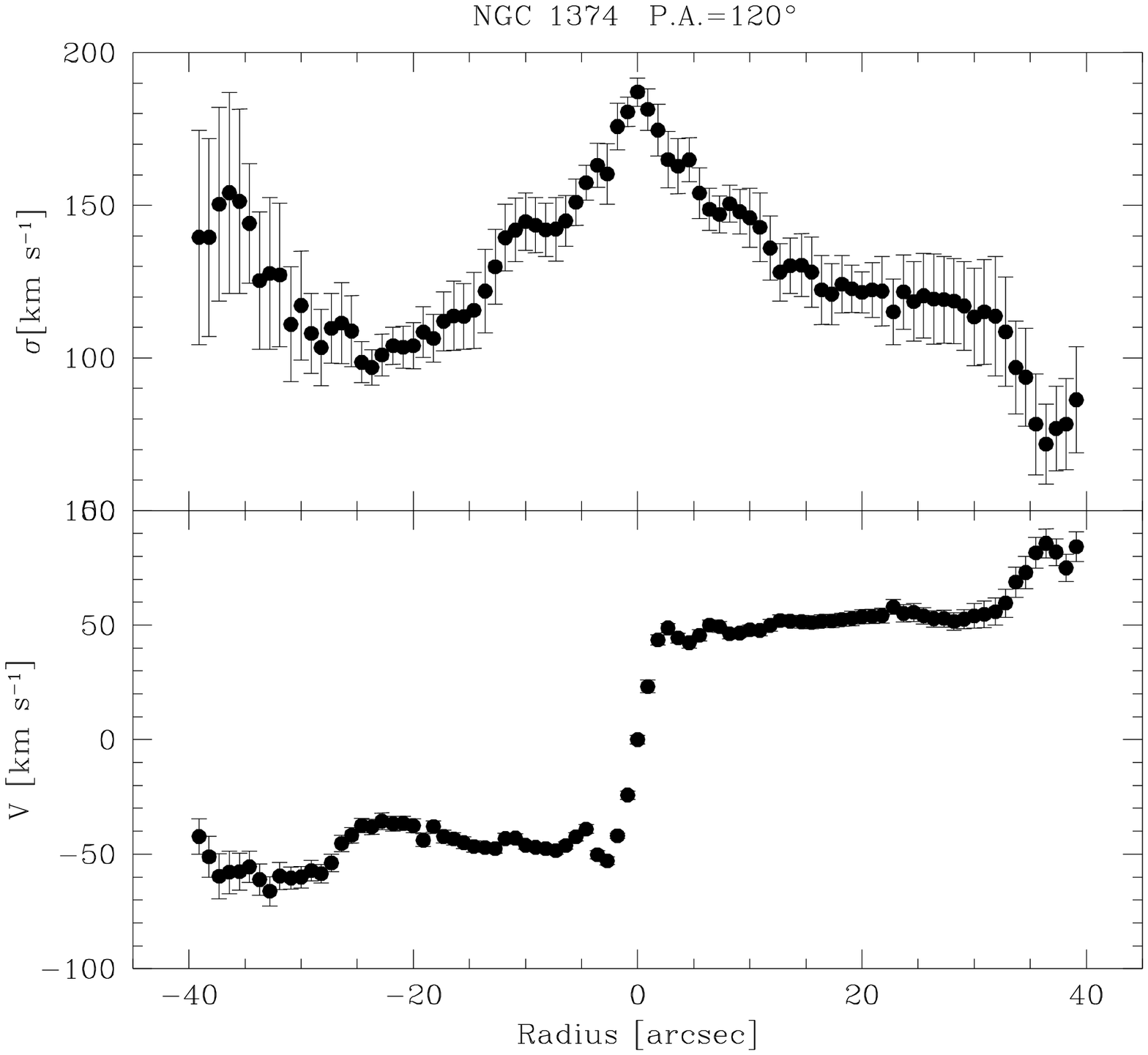}}
\caption{NGC~1374}
\label{n1374}
\end{figure}

\begin{figure}
\resizebox{\hsize}{!}{\includegraphics{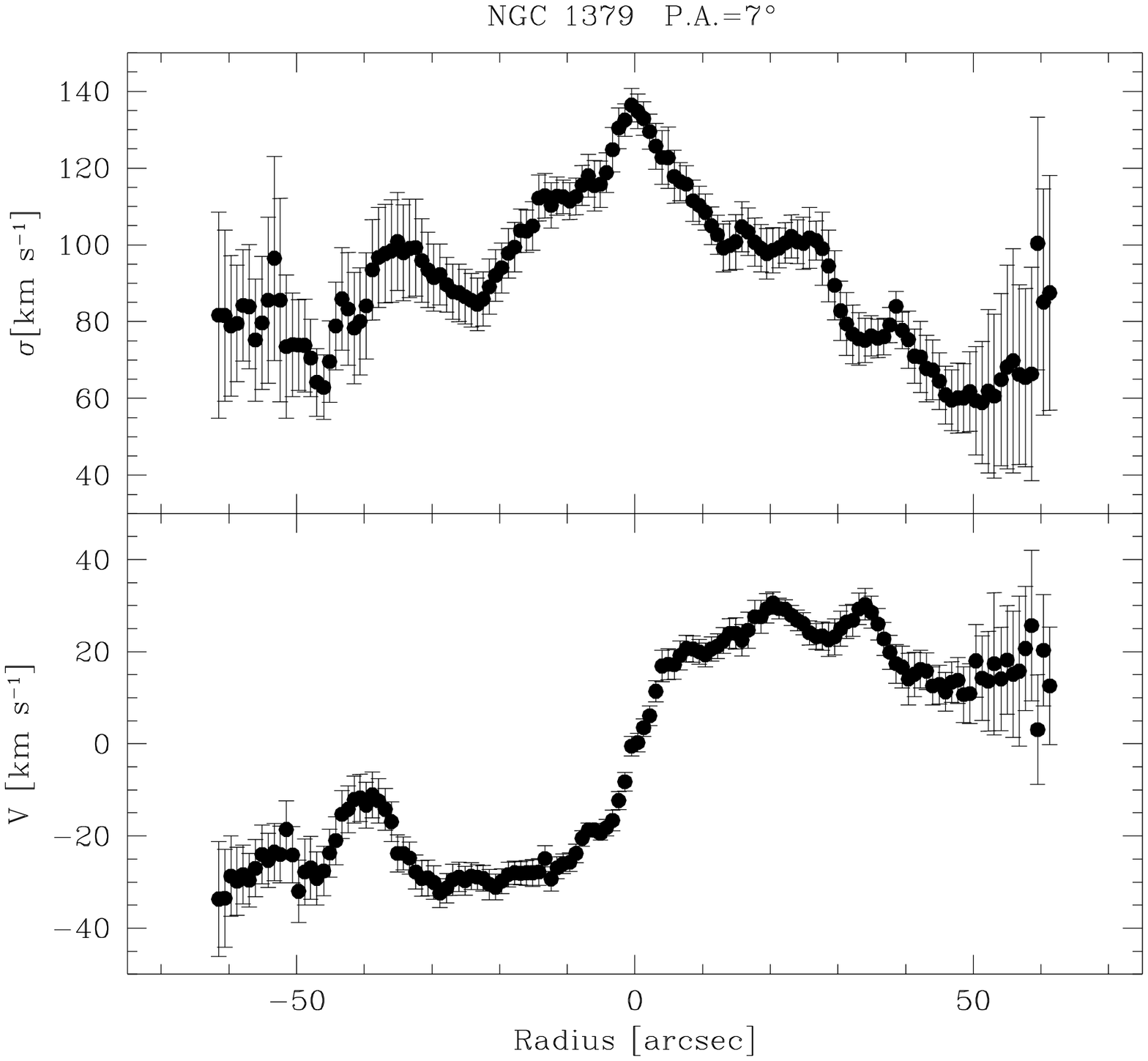}}
\caption{NGC~1379}
\label{n1379}
\end{figure}

%\pagebreak
%\newpage

\begin{figure}
\resizebox{\hsize}{!}{\includegraphics{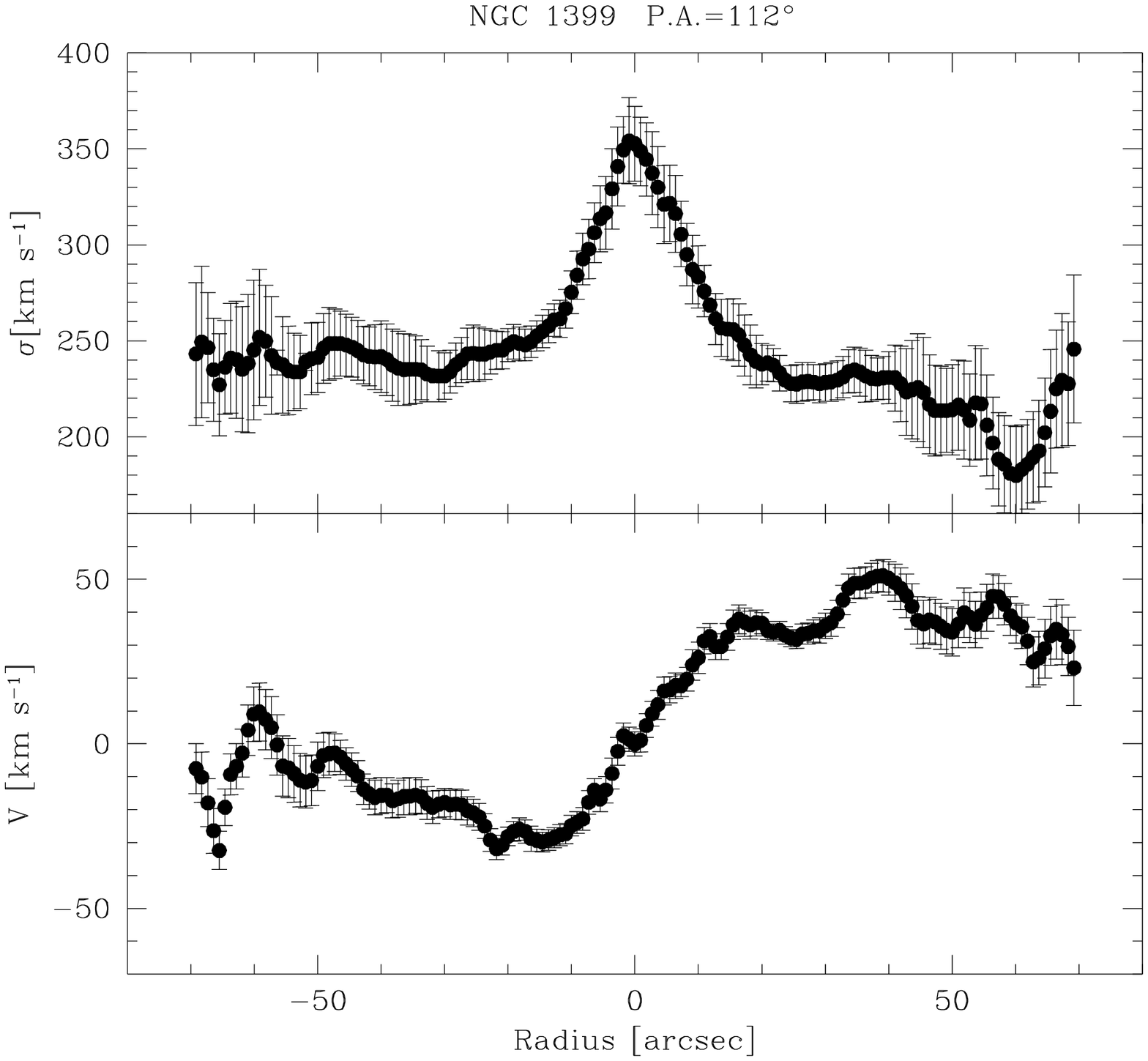}}
\caption{NGC~1399}
\label{n1399}
\end{figure}

\begin{figure}
\resizebox{\hsize}{!}{\includegraphics{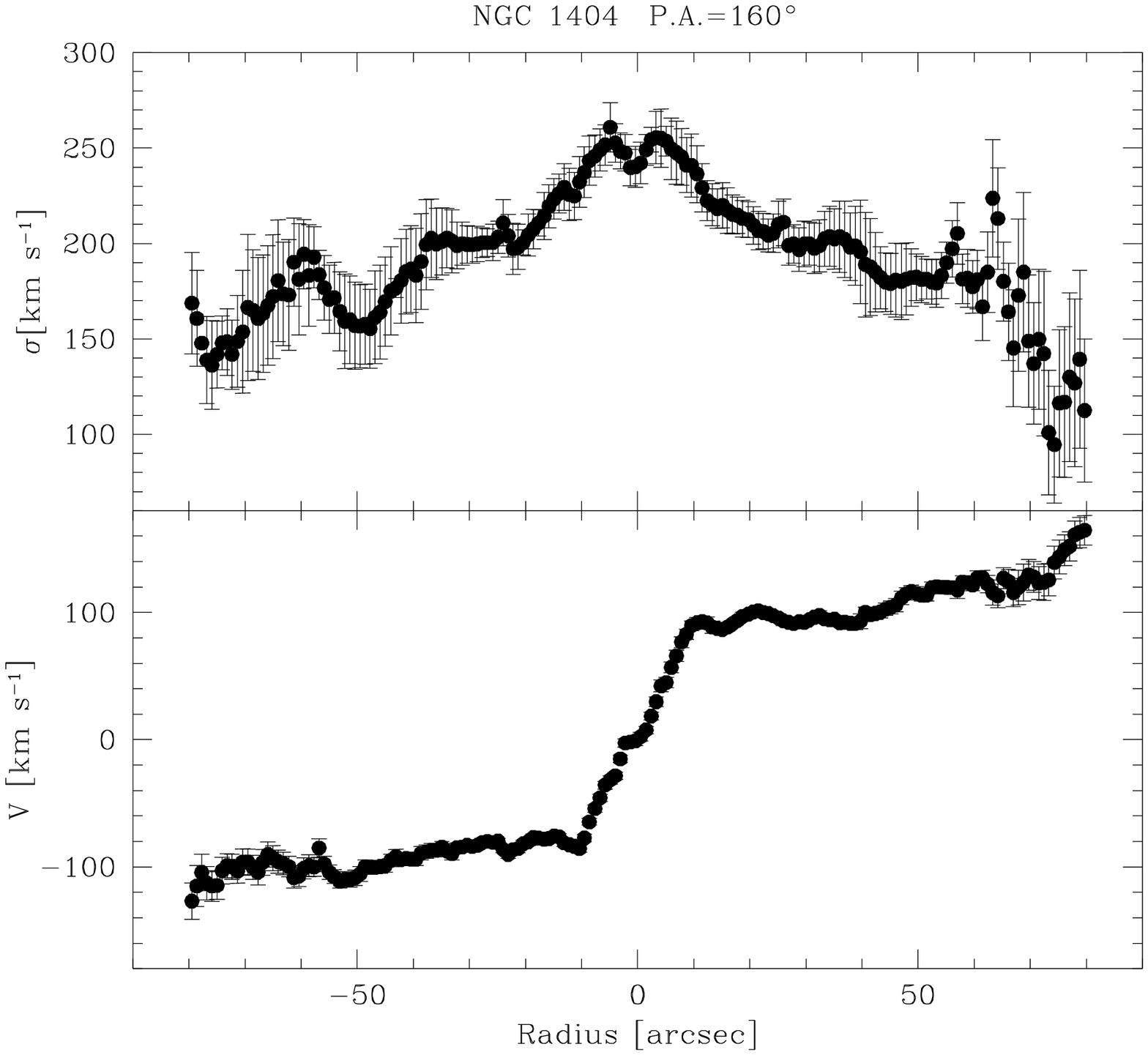}}
\caption{NGC~1404}
\label{n1404}
\end{figure}

\begin{figure}
\resizebox{\hsize}{!}{\includegraphics{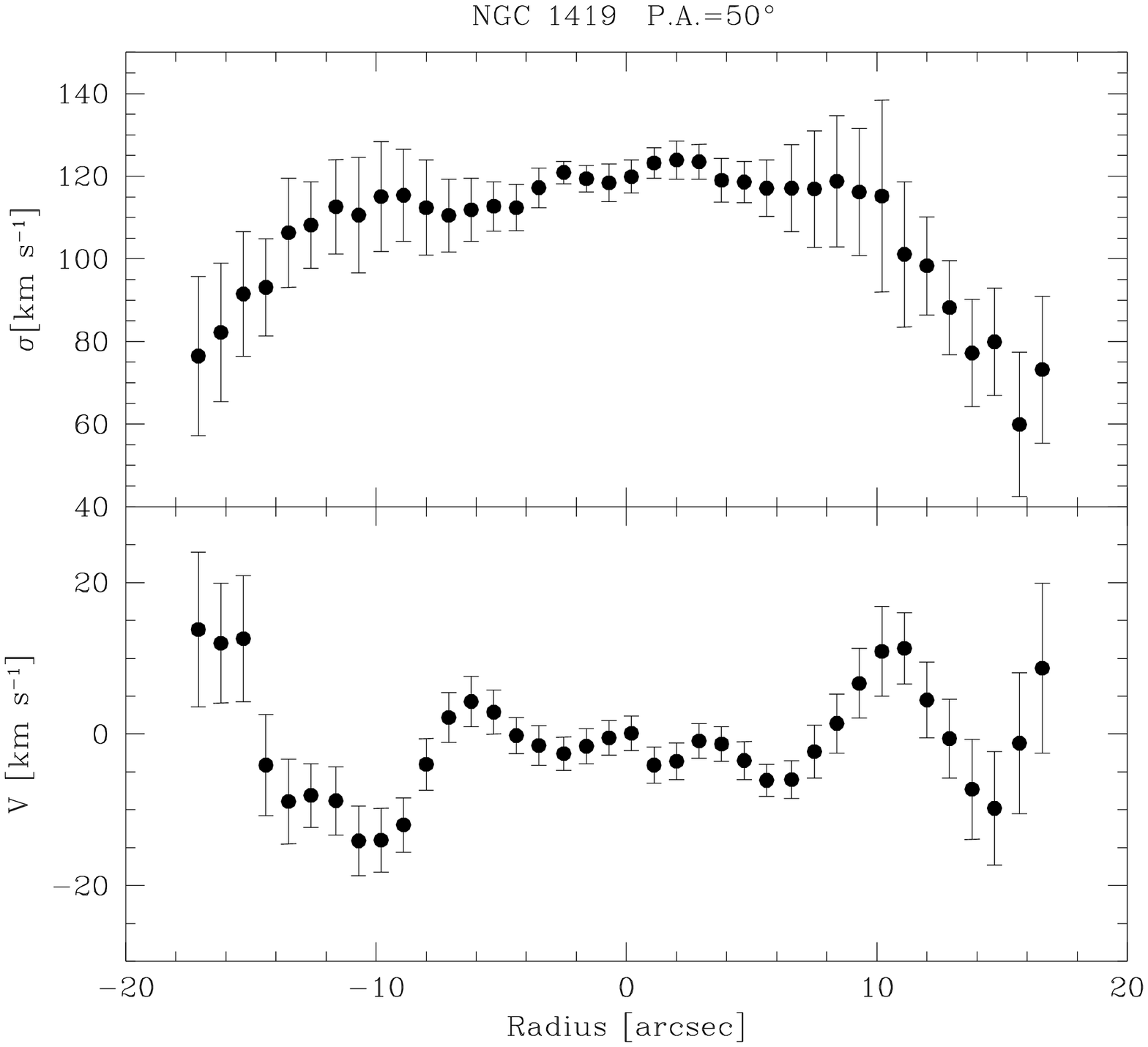}}
\caption{NGC~1419}
\label{n1419}
\end{figure}

\begin{figure}
\resizebox{\hsize}{!}{\includegraphics{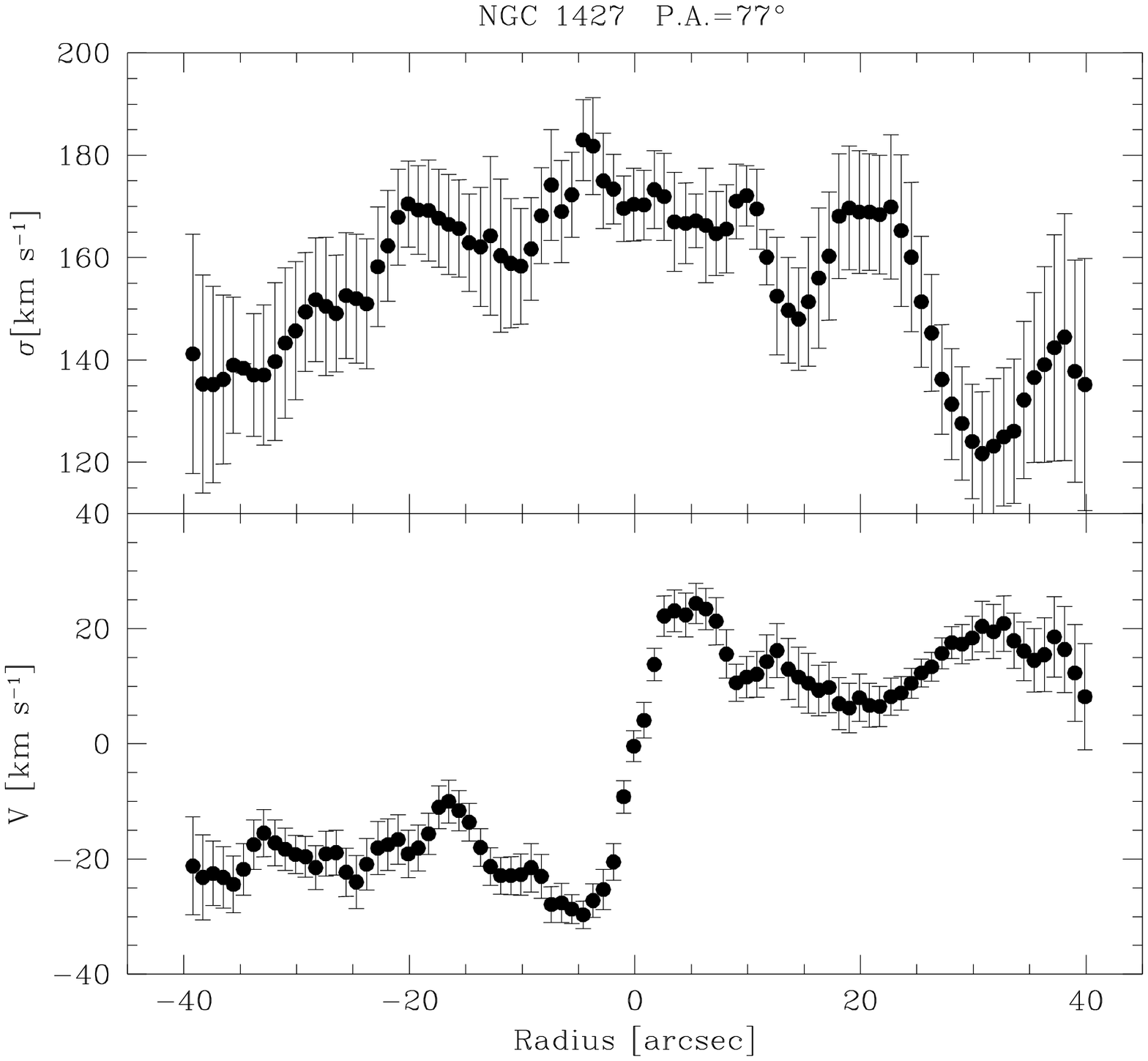}}
\caption{NGC~1427}
\label{n1427}
\end{figure}

\begin{figure}
\resizebox{\hsize}{!}{\includegraphics{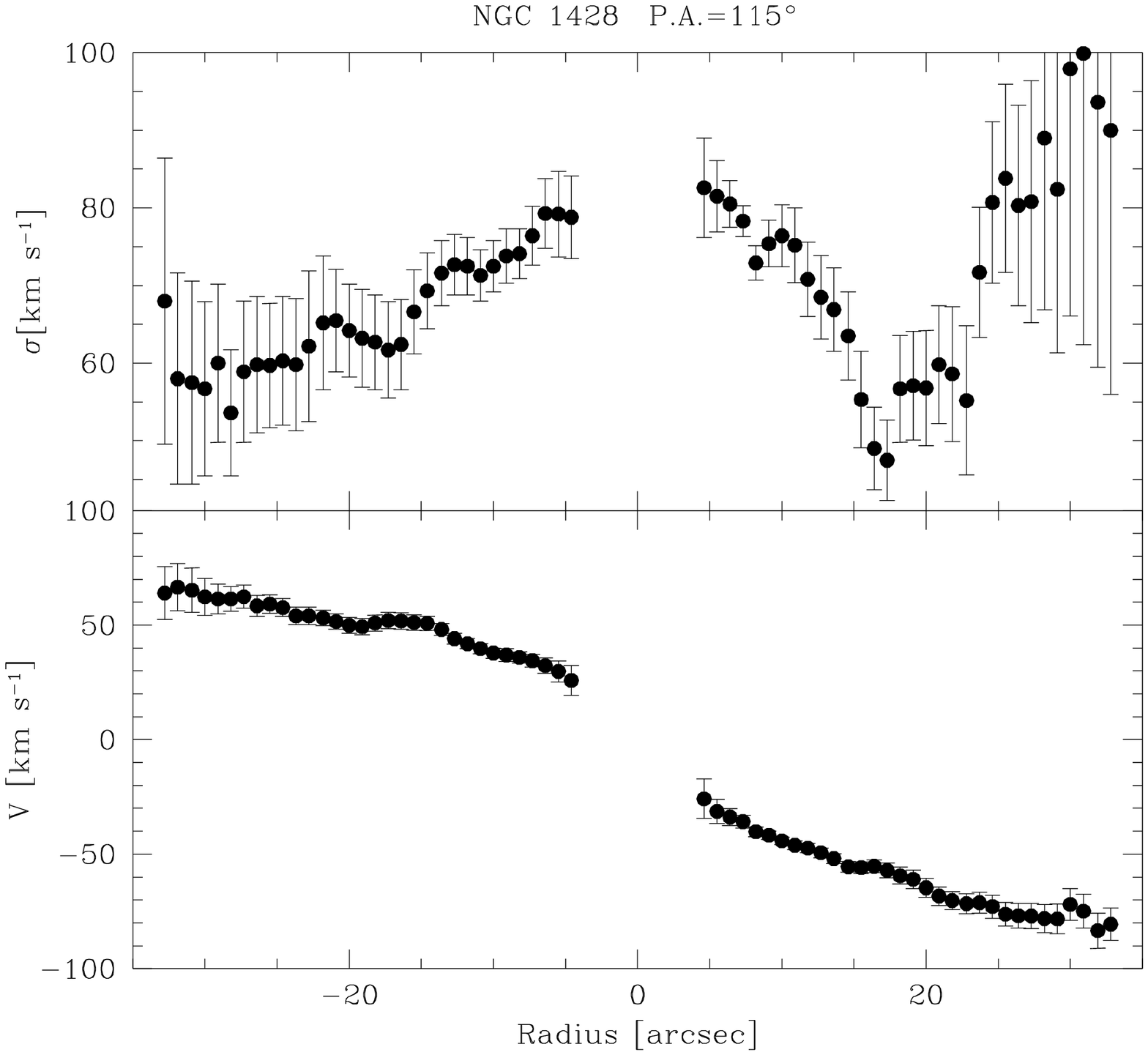}}
\caption{NGC~1428}
\label{n1428}
\end{figure}

\begin{figure}
\resizebox{\hsize}{!}{\includegraphics{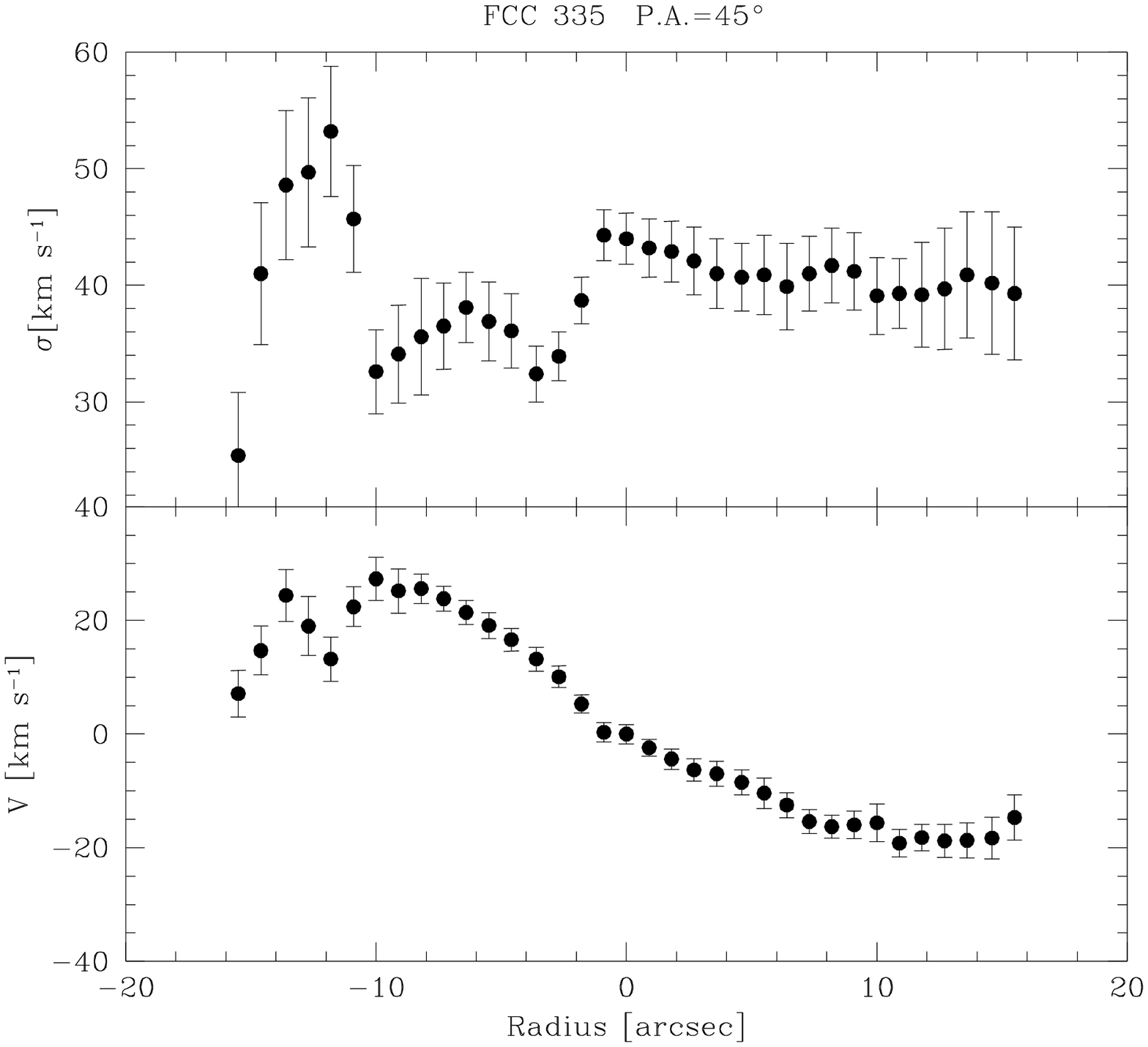}}
\caption{FCC~335}
\label{f335}
\end{figure}

\section{Literature comparison of kinematic profiles}

Comparison with previous data. 
Upper panels: velocity dispersion, lower panels: rotation curve.
Different sources are marked with different symbols as explained in the 
captions.
Notice that, in general, the P.A. of different authors do not exactly 
coincide. Those which are different from ours by more than 5$^{\circ}$ 
are the following: Bicknell et al. 1989 and Stiavelli et al. 1993 adopted 
P.A.=84$^{\circ}$ for NGC 1399 (instead of 112$^{\circ}$); 
van der Marel \& Franx 1993 adopted P.A.=110$^{\circ}$ for NGC 1374
(instead of 120$^{\circ}$); D95 adopted P.A.=65$^{\circ}$ for NGC 1419
(instead of 50$^{\circ}$).

\begin{figure*}
\resizebox{\hsize}{!}{\includegraphics{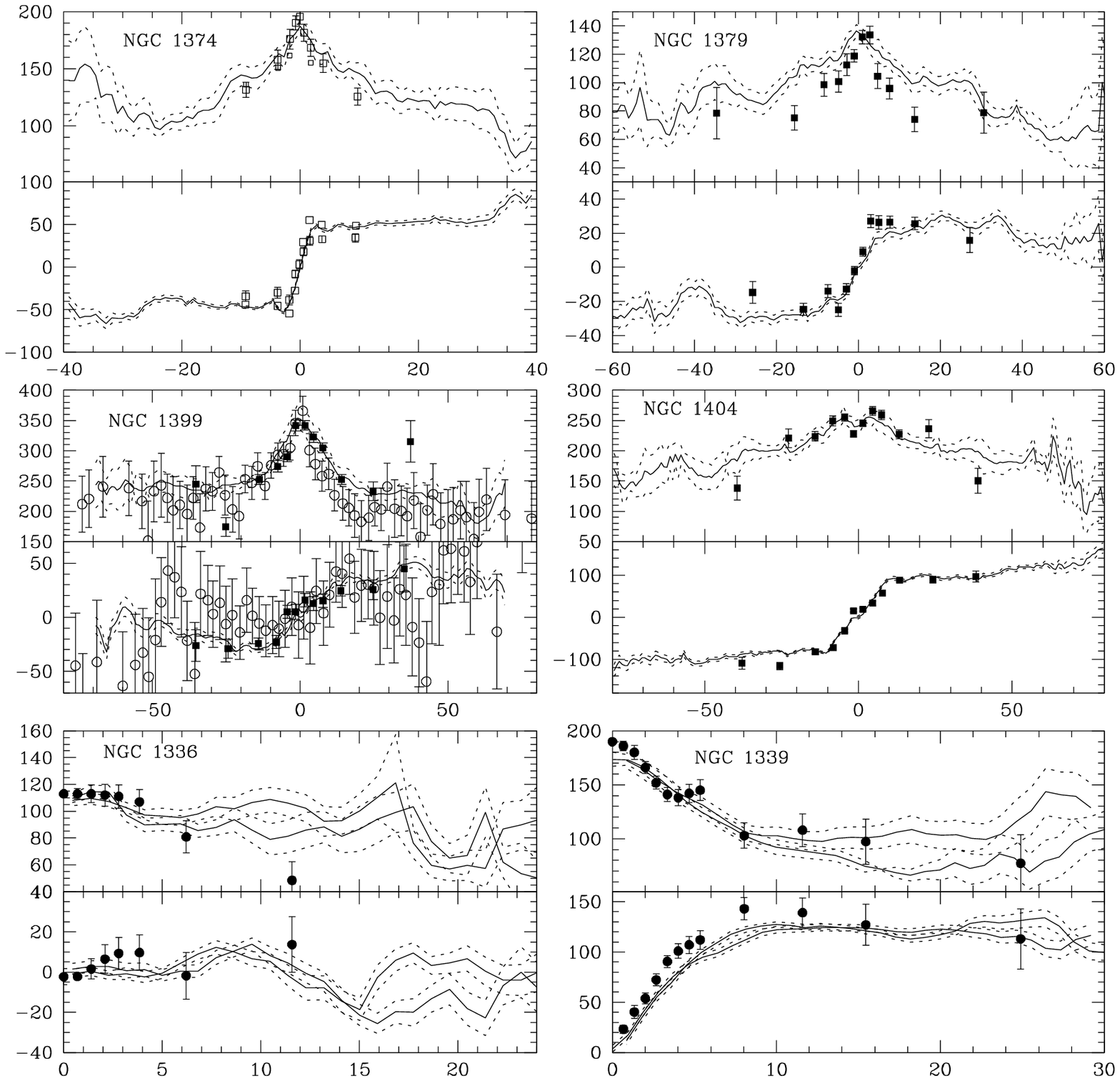}}
\caption{Comparison with previous data. Upper panels: velocity 
dispersion, lower panels: rotation curve. The radial scale is in 
arcseconds. The continuous lines represent the present data, while the 
errors are marked by the dashed lines. 
Open boxes: Van der Marel \& Franx 1993; filled boxes: Franx et al.
1989; open circles: Bicknell et al. 1989; filled dots: D'Onofrio et al.
1995.}
\label{compar1}
\end{figure*}

\begin{figure*}
\resizebox{\hsize}{!}{\includegraphics{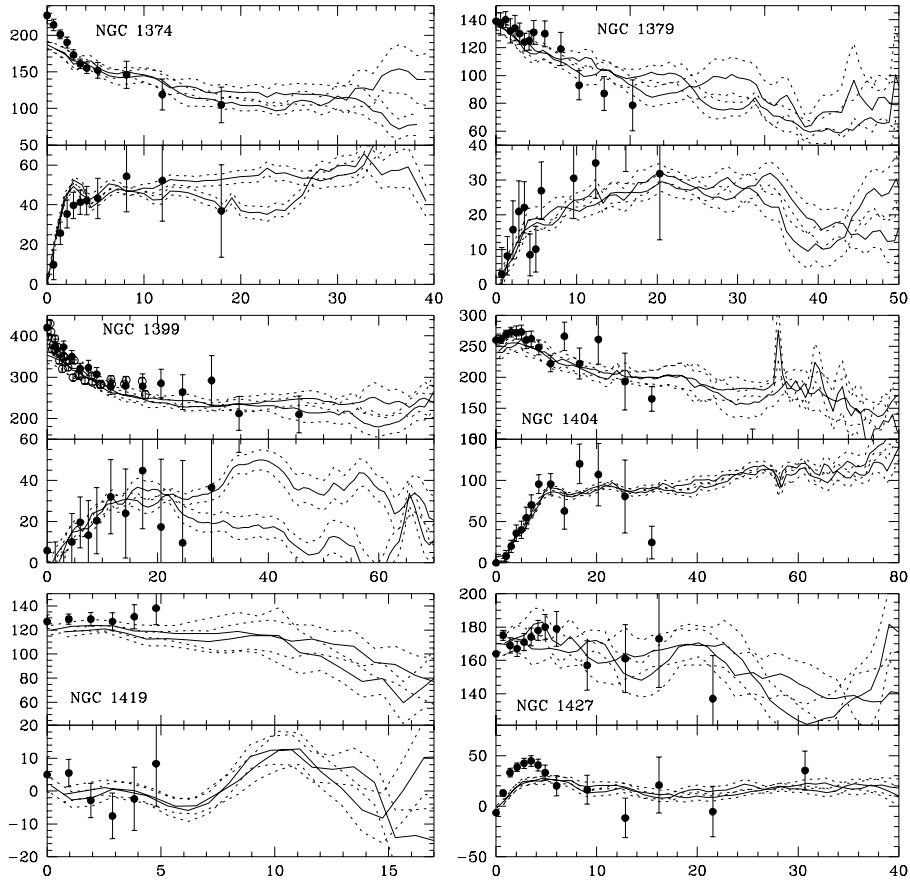}}
\caption{As in Fig.~\ref{compar1}. 
Filled circles: D'Onofrio et al. 1995; open circles: Stiavelli et al. 1993.}
\label{compar2}
\end{figure*}

\end{document}